\documentclass{llncs}
\usepackage{amsfonts}
\usepackage[dvips]{graphicx}
\usepackage{amsmath}
\usepackage{amssymb}
\usepackage{xspace}

\newlength{\defbaselineskip}
\newlength{\mylength}

\newcommand{\defAs}
{\:\mbox{\hbox{$= \! \! \raisebox{-0.5 ex}[0 ex][0 ex]{\tiny
Def}$}}\:}

\newcommand{\tb}[1]{\underline{\textbf{#1}}}

\newcommand{\adv}{\mathit{adv}}

\newcommand{\hor}{\textsf{Horspool}\xspace}
\newcommand{\quick}{\textsf{Quick-Search}\xspace}
\newcommand{\berry}{\textsf{Berry-Ravindran}\xspace}
\newcommand{\bm}{\textsf{Boyer-Moore}\xspace}
\newcommand{\fs}{\textsf{Fast-Search}\xspace}

\newcommand{\tbm}{\textsf{Tuned-Boyer-Moore}\xspace}

\newcommand{\smith}{\textsf{Smith}\xspace}

\title{On Tuning the Bad-Character Rule:\\the Worst-Character Rule}

\author{Domenico Cantone \and Simone Faro}
\institute{
Universit\`a di Catania, Dipartimento di Matematica e
Informatica\\Viale Andrea Doria 6, I-95125 Catania, Italy\\
\email{ cantone@dmi.unict.it,
faro@dmi.unict.it}
}

\begin{document}
\maketitle

\begin{abstract}
In this note we present the \emph{worst-character rule}, an efficient
variation of the \emph{bad-character} heuristic for the exact string
matching problem, firstly introduced in the well-known \bm
algorithm.
Our proposed rule selects a position relative to the current shift
which yields the largest average advancement, according to the
characters distribution in the text.
Experimental results show that the worst-character rule achieves very
good results especially in the case of long patterns or small
alphabets in random texts and in the case of texts in natural
languages.\\

\end{abstract}


\section{Introduction}
 Given a text $T$ and a pattern $P$ over some alphabet $\Sigma$, the
 \emph{string matching problem} consists in finding \emph{all}
 occurrences of the pattern $P$ in the text $T$. It is a very
 extensively studied problem in computer science, mainly due to its
 direct applications to such diverse areas as text, image and signal
 processing, information retrieval, computational biology, etc.

In this paper we present the \emph{worst-character rule}, an efficient
variation of the \emph{bad-character} heuristic for the exact string
matching problem, firstly introduced in the well-known \bm
algorithm~\cite{BM77}.
Our proposed rule selects a position relative to the current shift
which yields the largest average advancement, according to the
characters distribution in the text.
Experimental results show that the worst-character rule achieves very
good results especially in the case of long patterns or small
alphabets in random texts and in the case of texts in natural
languages.


Before entering into details, we review some useful notations and
terminology.  A string $P$ of length $m \geq 0$ over a finite alphabet
$\Sigma$ is represented as a finite array $P[0\,..\,m-1]$.  By $P[i]$
we denote the $(i+1)$-st character of $P$, for $0\leq i < m$.
Likewise, by $P[i\,..\,j]$ we denote the substring of $P$ contained
between the $(i+1)$-st and the $(j+1)$-st characters of $P$, where
$0\leq i \leq j < m$.

Let $T$ be a text of length $n$ and let $P$ be a pattern of length
$m$.  If the character $P[0]$ is aligned with the character $T[s]$
of the text, so that $P[i]$ is aligned with $T[s+i]$, for
$0 \leq i \leq m-1$, we say that the pattern $P$ has \emph{shift} $s$
in $T$.  In this case the substring $T[s\,..\, s+m-1]$ is called the
\emph{current window} of the text.  If $T[s\,..\, s+m-1] = P$, we
say that the shift $s$ is \emph{valid}.
Then the \emph{string matching problem} consists in finding all valid
shifts of $P$ in $T$, for given pattern $P$ and text $T$.

\begin{figure}[!t]
\renewcommand{\baselinestretch}{1.1}
\begin{center}
    \begin{scriptsize}
        \begin{tabular}{|ll|ll|}
        \hline
&&&\\
        ~~\textbf{(A)}&{\textsc{Generic\_String\_Matcher($T$, $P$, $n$,
$m$)}}& ~~\textbf{(B)}& \textsc{Precompute\_bc}$(P, \Sigma)$ \\
        ~~~1. & \qquad \textsc{Precompute\_Globals}$(P)$&~~~1. &\qquad $m
=$ length($P$) \\
        ~~~2. & \qquad $s := 0$&~~~2. &\qquad \textbf{for each} $c \in
\Sigma$ \textbf{do} \\
        ~~~3. & \qquad \textbf{while} $s\leq n-m$ \textbf{do}&~~~3.
&\qquad \qquad $bc_{P}(c)=m$ \\
        ~~~4. & \qquad \qquad $j :=$ \textsc{Check\_Shift}$(s,P,T)$&~~~4.
&\qquad \textbf{for} $i = 0$ \textbf{to} $m-1$ \textbf{do} \\
        ~~~5. & \qquad \qquad $s := s +$
\textsc{Shift\_Increment}$(s,P,T,j)$~~~&~~~5. &\qquad \qquad
$bc_{P}(P[i])=m-i-1$ ~~~\\
&&&\\
        \hline
        \end{tabular}
    \caption{\textbf{(A)} The procedure
\textsc{Generic\_String\_Matcher} for searching the occurrences of a
pattern
    $P$ in a text $T$. \textbf{(B)} The procedure
\textsc{Precompute\_bc} for computing the bad-character heuristic.}
    \label{fig:bad_character}
    \end{scriptsize}
\end{center}
\renewcommand{\baselinestretch}{1.5}
\end{figure}

Most string matching algorithms have the general structure shown in
Figure~\ref{fig:bad_character}(A), where the procedure
$\textsc{Precompute\_Globals}(P)$ computes useful mappings, in the
form of tables, which may be accessed by the function named
$\textsc{Shift\_Increment}(s,P,T,j)$; the function
$\textsc{Check\_Shift}(s,P,T)$ checks whether $s$ is a valid shift and
returns the position $j$ of the last matched character in the pattern;
the function $\textsc{Shift\_Increment}(s,P,T,j)$ computes a
\emph{positive} shift increment according to the information tabulated
by procedure named $\textsc{Precompute\_Globals}(P)$ and to the position $j$
of the last matched character in the pattern.
For instance, to look for valid shifts, the
celebrated \bm algorithm \cite{BM77} scans the pattern from
right to left and, at the end of the matching phase, it computes the
shift increment as the largest value given by the \emph{good-suffix}
and the \emph{bad-character} rules.

\section{The bad-character rule}\label{sec:preprocessing}
Information gathered during the execution of the
$\textsc{Shift\_Increment}(s,P,T,j)$ function, in combination with
the knowledge of $P$, as suitably extracted by procedure
$\textsc{Precompute\_Globals}(P)$, can yield shift increments larger
than $1$ and ultimately lead to more efficient algorithms. In this
section we focus our attention on the use of the bad-character
heuristic for preprocessing the pattern, introduced by Boyer and
Moore in~\cite{BM77}.

The \bm algorithm is the progenitor of several algorithmic variants
which aim at computing close to optimal shift increments very efficiently.
 Specifically, the \bm algorithm checks whether $s$ is a valid shift,
 by scanning the pattern $P$ from right to left and, at the end of the
 matching phase, it computes the shift increment as the largest value
 suggested by the \emph{good-suffix rule} and the \emph{bad-character
 rule}, provided that both of them are applicable.

Specifically, the bad-character heuristic states that if
$c=T[s+j-1]\neq P[j-1]$ is the first mismatching character, while
scanning $P$ and $T$ from right to left with shift $s$, then $P$ can
be safely shifted in such a way that its rightmost occurrence of $c$,
if present, is aligned with position $(s+j-1)$ in $T$ (provided that
such an occurrence is in $P[0\,..\,j-2]$, otherwise the bad-character
rule has no effect).
In the case in which $c$ does not occur in $P$, then $P$ can be safely
shifted just past position $(s+j-1)$ in $T$.  More formally, the shift
increment suggested by the bad-character heuristic is given by the
expression $(j-bc_{_P}(T[s+j-1])-1)$, where $ bc_{_P}(c) \defAs \max
(\{0\leq k < m \:|\: P[k] = c\} \cup \{-1\})\,$,  for $c \in
\Sigma$.
Procedure \textsc{Precompute\_bc}, shown in
Figure~\ref{fig:bad_character}(B), computes the function $bc_{_P}$
during the preprocessing phase in $\mathcal{O}(m + \sigma)$-time and
$\mathcal{O}(\sigma)$-space, where $\sigma$ is the size of the
alphabet $\Sigma$.

Due to the simplicity and ease of implementation of the
bad-character heuristic, some variants of the \bm algorithm were
based just on it and dropped the good-suffix heuristic.

For instance, Horspool~\cite{Hor80} suggested the following
simplification of the original \bm algorithm, which performs better in
practical cases.  He just dropped the good suffix heuristic and
proposed to compute shift advancements in such a way that the
rightmost character $T[s+m-1]$ is aligned with its rightmost
occurrence on $P[0\,..\,m-2]$, if present; otherwise the pattern is
advanced just past the window.  This corresponds to advance the shift
by $hbc_P(T[s+m-1])$ positions, where
$$ hbc_{_P}(c) \defAs \min
(\{1\leq k < m \:|\: P[m-1-k] = c\} \cup\{m\}) \enspace.  
$$

The resulting algorithm performs well in practice and can be
immediately translated into programming code (see Baeza-Yates and
R{\'e}gnier~\cite{BYR92} for a simple implementation in the
\textbf{C} programming language).

Likewise, the \quick algorithm, presented in~\cite{Sun90}, uses a
modification of the original heuristic, much along the same lines of
the \hor algorithm.  Specifically, it is based on the following
observation: when a mismatching character is encountered, the pattern is
always shifted to the right by at least one character, but never by
more than $m$ characters.  Thus, the character $T[s+m]$ is always
involved in testing for the next alignment.  So, one can apply the bad
character rule to $T[s+m]$, rather than to the mismatching character,
possibly obtaining larger shift advancements.  This corresponds to
advance the shift by $qbc_P(T[s+m])$ positions, where
$$ 
qbc_{_P}(c)
\defAs \min (\{1 \leq k \leq m \:|\: P[m-k] = c\} \cup\{m+1\})\enspace.  
$$

Finally, the \smith algorithm~\cite{Smi91} computes its shift
advancements by taking the largest value suggested by the \hor and the \quick
bad-character rules.
Its preprocessing phase is
performed in $O(m+\sigma)$-time and $O(\sigma)$-space complexity,
while its
searching phase has a quadratic worst case time.

Although the role of the good-suffix heuristic in practical string matching algorithms
has recently been reappraised \cite{CF03a,CF03b,CF05}, also in consideration of the fact that often it is
as effective as the bad-character heuristic, especially in the case of non-periodic
patterns, the bad character heuristic is still considered one of the powerfull method for
speed up the performance of string matching algorithms (see for instance \cite{FL08,FL09}).

\section{The worst-character rule}
\label{sec:tuning}

For a given shift $s$, the \hor and the \quick algorithms compute
their shift advancements by applying the bad-character rule on a fixed
position $s+q$ of the text, with $q$ equal respectively to $m-1$ and
to $m$.  We refer to the value $q$ as the \emph{bad-character relative
position}.

It may be possible that other bad-character relative
positions generate larger shift advancements. We will show below
how, given a pattern $P$ and a text $T$ with known character
distribution, we can compute efficiently the bad-character relative
position, to be called \emph{worst-character relative position},
which ensures the largest shift advancements on the average.
The \emph{worst-character rule} is then the bad-character rule based
on such a worst-character relative position.


\subsection{Finding the worst-character relative position}
To begin with, we introduce the \emph{generalized bad-character function}
$gbc_{_P}(i,c)$.
%
%
Suppose the pattern $P$ has shift $s$ in the text $T$.  For a given
bad-character relative position $i$, with $0\leq i\leq m$,
$gbc_{_P}(i,T[s+i])$ is the shift advancement such that the character
$T[s+i]$ is aligned with its rightmost occurrence in $P[0\,..\,i-1]$,
if present; otherwise $gbc_{_P}(i,T[s+i])$ evaluates to $i+1$ (this
corresponds to advance the pattern just past position $s+i$ of the
text). Thus,
%
%
%
$$
gbc_{_P}(i,c) \defAs \min (\{1\leq k \leq i \:|\: P[i-k] = c\}
\cup\{i+1\}), \enspace\ \textrm{ for } c \in \Sigma , ~
0 \leq i \leq m\,.
$$
Plainly, $gbc_{_P}(i,c) \geq 1$ always holds. Additionally,
the shift rules of the \hor and \quick algorithms
can be expressed in terms of the generalized bad-character function
by $hbc_{_P}(c)=gbc_{_P}(m-1,c)$ and $qbc_{_P}(c)=gbc_{_P}(m,c)$,
respectively, for $c \in \Sigma$.


Next, let $f:\Sigma\rightarrow [0,1]$ be the relative frequency of the
characters in the text $T$.  Given a fixed pattern $P$ and a
bad-character relative position $0 \leq i \leq m$, the average shift
advancement of the generalized bad-character function on $i$ is given
by the function
$$
\adv_{_{P,f}}(i)=_{\mathit{Def}} \sum_{c\in \Sigma} f(c) \cdot gbc_{_P}(i,c)\,.
$$

Thus, the \emph{worst-character relative position} of a given pattern
$P$ and a given relative frequency function $f$ can be defined as the
smallest position $0 \leq q \leq m$ such that
$$
     \adv_{_{P,f}}(q) = \max_{0 \leq j \leq m} \adv_{_{P,f}}(j)\,.
$$
The procedure \textsc{Find\_worst\_character}, shown in
Figure~\ref{fig:worst_character}(A), computes the worst-character
relative position for a given input pattern $P$ and a given relative
frequency function $f$ over $\Sigma$ in $\mathit{O}(m+\sigma)$-time
and $\mathit{O}(\sigma)$-space.
It exploits the recurrence
$$
\adv_{_{P,f}}(i) =\left\{
\begin{array}{ll}
       1 &  \mbox{if } i=0 \\
       \adv_{_{P,f}}(i-1)+1-f(P[i-1])\cdot gbc_{_P}(i-1,P[i-1])~~~~  &
       \mbox{if } 1\leq i\leq m
\end{array}
\right.
$$
for the computation of $\adv_{_{P,f}}(i)$, for $i=0,\ldots,m$, which
is based, in turn, on the fact that
$$
gbc_{_P}(i,c) = \left\{
\begin{array}{ll}
     gbc_{_P}(i-1,c) & \mbox{if } P[i-1] \neq c\\
     1 & \mbox{otherwise\,,}
\end{array}
\right.
$$
for $c \in \Sigma$ and $i=0,1,\ldots,m$.

Observe that in the above recurrence only entries of the generalized
bad-character function of the form $gbc_{_P}(i,P[i])$ are needed.  To
compute such values, the characters of the pattern are processed from
left to right and, for each position $i$, the \emph{last position}
function $lp^i_{_P}:\Sigma \rightarrow\{-1,0,\ldots m-1\}$, which
gives the rightmost occurrence of each character $c \in \Sigma$ in
$P[0\,..\,i-1]$, is also computed.
The value of $lp^i_{_P}(c)$ is set to $-1$ if either $i=0$ or $c$ is not
present in $P[0\,..\,i-1]$.  Formally, for $c \in \Sigma$,
$$
lp^i_{_P}(c) = \max(\{0\leq j< i\ |\ P[j]=c \} \cup\{-1\}).
$$
Observe that at the $i$-th iteration of the \textbf{for}-loop of
procedure \textsc{Find\_worst\_character}, only the value
$lp^{i}_{_P}(P[i])$ is needed.  The function $lp^{i}_{_P}$ is
maintained as an array of dimension $\sigma$ and computed by the
following recursive relation
$$
lp^i_{_P}(c)=\left\{
\begin{array}{ll}
       -1  &  \mbox{if } i=0 \\
       i-1 &  \mbox{if } i>0 \mbox{ and } c = P[i-1]\\
       lp^{i-1}_{_P}(c) &  \mbox{if } i>0 \mbox{ and } c\neq P[i-1].
\end{array}
\right.
$$
The initialization of $lp^{0}_{_P}$ is plainly done in
$\mathit{O}(\sigma)$-time, while the computation of $lp^{i}_{_P}$, for
$i>0$, can be done in constant time from array $lp^{i-1}_{_P}$.
Finally, the values $gbc_{_P}(i,P[i])$ are computed using the
following relation
$$
gbc_{_P}(i,P[i])=\left\{
\begin{array}{ll}
       1 &  \mbox{if } i=0 \\
       i - lp^i_{_P}(P[i]) &  \mbox{if } 0 < i< m.
\end{array}
\right.
$$

\subsection{The worst-character heuristic}
The position $q$ computed by procedure
\textsc{Find\_worst\_character} is then used by the worst-character
heuristic to calculate shift advancements during the searching
phase. In particular the worst-character heuristic computes shift
advancements in such a way that the character $T[s+q]$ is
aligned with its rightmost occurrence on $P[0\,..\,q-1]$, if
present; otherwise the pattern is advanced just past position $s+q$
of the text. This corresponds to advance the shift by
$wc_{_P}(T[s+q])$ positions, where
$$
wc_{_P}(c) \defAs \min (\{1\leq k \leq q \:|\: P[q-k] = c\}
\cup\{q+1\}) \enspace.
$$
Observe that if $q=0$ then the advancement is always equal to 1. The
resulting algorithm can be immediately translated into programming
code (see Figure~\ref{fig:worst_character}(C) for a simple
implementation). The procedure \textsc{Precompute\_wc}, shown in
Figure~\ref{fig:worst_character}(B), computes the
table which implements the worst-character heuristic in
$\mathit{O}(m+\sigma)$-time and space.

\begin{figure}[!t]
\renewcommand{\baselinestretch}{1.1}
    \begin{center}
    \begin{scriptsize}
    \begin{tabular}{|ll|ll|}
    \hline
&&&\\
    ~~~\textbf{(A)}&\textsc{Find\_worst\_character$(P, \Sigma, f)$}
&~~~\textbf{(B)}&\textsc{Precompute\_wc$(P, \Sigma, q)$}\\
    ~~~~1. &\quad $m =$ length($P$)
&~~~~1. &\quad $m =$ length($P$)\\
    ~~~~2. &\quad \textbf{for each } $c \in \Sigma$ \textbf{do}
&~~~~2. &\quad \textbf{for each} $c \in \Sigma$ \textbf{do}\\
    ~~~~3. &\quad \quad $lp_{_P}(c)=-1$
&~~~~3. &\quad \quad $wc(c)=q+1$\\
    ~~~~4. &\quad $q = 0$
&~~~~4. &\quad \textbf{for} $i = 0$ \textbf{to} $q-1$ \textbf{do}\\
    ~~~~5. &\quad $\adv_{_{P,f}}(0) = 1$
&~~~~5. &\quad \quad $wc(P[i])=q-i$\\
    ~~~~6. &\quad $max = 1$                                            &&\\
    ~~~~7. &\quad $lp_{_P}(P[0]) = 0$
&~~~\textbf{(C)}&\textsc{Worst\_Character\_Matcher($P,T,m,n$)}\\
    ~~~~8. &\quad $\delta = f(P[0])$
&~~~~1. &\quad $q = $\textsc{Find\_worst\_character$(P, \Sigma, f)$}\\
    ~~~~9. &\quad \textbf{for} $i = 1$ \textbf{to} $m$ \textbf{do}
&~~~~2. &\quad $wc$ = \textsc{Precompute\_wc$(P, \Sigma, q)$}\\
    ~~~10. &\quad \quad $\adv_{_{P,f}}(i) = \adv_{_{P,f}}(i-1) + 1 -
\delta$~~~    &~~~~3. &\quad $s = 0$\\
    ~~~11. &\quad \quad $\delta = f(P[i]) \cdot (i-lp_{_P}(P[i]))$
&~~~~4. &\quad \textbf{while} $s \leq n - m$ \textbf{do}\\
    ~~~12. &\quad \quad $lp_{_P}(P[i])=i$
&~~~~5. &\quad \quad $j=m-1$\\
    ~~~13. &\quad \quad \textbf{if} $\adv_{_{P,f}}(i) > max$
\textbf{then} &~~~~6. &\quad \quad \textbf{while} $j\geq 0$
\textbf{and} $P[j]=T[s+j]$ \textbf{do}~~~\\
    ~~~14. &\quad \quad \quad $max = \adv_{_{P,f}}(i)$
&~~~~7. &\quad \quad \quad $j=j-1$\\
    ~~~15. &\quad \quad \quad $q = i$
&~~~~8. &\quad \quad \textbf{if} $j<0$ \textbf{then}
\textsc{Output}($s$)\\
    ~~~16. &\quad \textbf{return} $q$
&~~~~9. &\quad \qquad $s = s+wc(T[s+q])$\\
&&&\\
    \hline
    \end{tabular}
    \caption{\textbf{(A)} The procedure \textsc{Find\_worst\_character}
    for computing the worst-character relative position of the pattern
    $P$.  \textbf{(B)} The procedure \textsc{Precompute\_wc} for
    computing the worst-character heuristic \textbf{(C)} The
    \textsc{Worst\_Character\_Matcher} algorithm which makes use of the
    worst-character heuristic.}
    \label{fig:worst_character}
    \end{scriptsize}
    \end{center}
\renewcommand{\baselinestretch}{1.5}
\end{figure}

\section{Experimental Results}\label{sec:experimental}

To evaluate experimentally the impact of the worst-character
heuristic, we have chosen to test the
\textsc{Worst\_Character\_Matcher} algorithm (in short \textsf{WC}),
given in Figure~\ref{fig:worst_character}(C), with three algorithms
based on variations of the bad-character rule, namely the \hor
algorithm (in short, \textsf{HOR}), the \quick algorithm (in short
\textsf{QS}), and the \smith algorithm (in short \textsf{SM}).
Experimental results have been evaluated in terms of running times
 and average advancement given by the shift heuristics.
All algorithms have been implemented in the \textsf{C} programming
language and were used to search for the same strings in large fixed
text buffers on a PC with AMD Athlon processor of 1.19GHz.  In
particular, all algorithms have been tested on four
\textsf{Rand}$\sigma$ problems and on four \textsf{Exp}$^{\lambda}\sigma$
problems, for alphabet sizes $\sigma=2,4,8,16$.  For each problem, the
patterns have been constructed by selecting 200
random substrings of length $m$ from the files, for
$m=2,4,8,16,32,64,128,256,512$.

Each \textsf{Rand}$\sigma$ and \textsf{Exp}$^\lambda\sigma$ problem
consists in searching a set of $200$ random patterns of a given
length in a 20Mb random text over a common alphabet of size
$\sigma$. \textsf{Rand}$\sigma$ and \textsf{Exp}$^\lambda\sigma$ problems
differ in the distribution of characters in the text buffer.

In a \textsf{Rand}$\sigma$ the characters of the text buffer have a
uniform distribution, i.e. the relative characters frequency is
defined by the law $f(c) = 1/\sigma$,  for all $c \in \Sigma$.

In an \textsf{Exp}$^\lambda\sigma$ problem the distribution of
characters follows the inverse-rank power-law of degree $\lambda$, a
model that gives a very good approximation of the relative frequency
function of characters in terms of their ranks both in natural language
dictionaries and texts (cf.\ \cite{CF03c}).  Formally, in a text in
natural language the relative frequency of the character $c_i$ of
rank $i$ can be approximated by
$$
f(c_i) = \frac{(\sigma - i + 1)^\lambda}{\sum_{j=1}^{\sigma}j^\lambda},
\hbox{~~for } i = 1,\ldots,\sigma\,,
$$
where the value of the degree $\lambda \in \mathbb{R}$ can be
determined experimentally and usually ranges in the interval $[3..10]$
(cf.\ \cite{CF03c}).  In our tests we have set $\lambda=5$.

In the following tables running times are expressed in hundredths of
seconds, while the average advancements are expressed in number of characters.

\renewcommand{\baselinestretch}{1.0}

\begin{scriptsize}
\begin{center}
\begin{tabular*}{0.95\textwidth}{@{\extracolsep{\fill}}|c|cccccccc|}
      \hline
      $\ \sigma=2$ & 2 & 4 & 8 & 16 & 32 & 64 & 128 & 256\\
      \hline
      \textsf{HOR}  & 47.78 & 47.55 & 46.70 & 47.90 & 44.49 & 44.42 &
44.79 & 44.35\\
      \textsf{QS}   & \tb{40.07} & 45.15 & 44.56 & 45.13 & 42.00 &
41.36 & 41.48 & 41.29\\
      \textsf{SM}   & 60.74 & 59.58 & 58.65 & 61.98 & 60.48 & 60.33 &
60.80 & 60.51\\
      \textsf{WC}   & 41.36 & \tb{43.78} & \tb{37.76} & \tb{33.55} &
\tb{28.08} & \tb{25.59} & \tb{23.94} & \tb{22.53}\\
      \hline
      $\ \sigma=4$ & 2 & 4 & 8 & 16 & 32 & 64 & 128 & 256\\
      \hline
      \textsf{HOR}  & 37.85 & 28.78 & 23.17 & 22.20 & 22.29 & 21.84 &
21.64 & 21.82\\
      \textsf{QS}   & \tb{29.97} & \tb{25.63} & 22.25 & 20.91 & 21.16
& 21.00 & 20.87 & 20.97\\
      \textsf{SM}   & 48.24 & 38.77 & 30.54 & 29.09 & 29.55 & 28.70 &
28.33 & 28.72\\
      \textsf{WC}   & 30.88 & 26.61 & \tb{22.22} & \tb{20.01} &
\tb{18.95} & \tb{18.36} & \tb{17.96} & \tb{17.45}\\
      \hline
      $\ \sigma=8$ & 2 & 4 & 8 & 16 & 32 & 64 & 128 & 256\\
      \hline
      \textsf{HOR}  & 30.01 & 22.15 & 18.55 & 17.33 & 17.07 & 17.00 &
16.95 & 17.05\\
      \textsf{QS}   & \tb{23.54} & \tb{20.15} & \tb{17.80} &
\tb{16.96} & 16.77 & 16.79 & 16.73 & 16.73\\
      \textsf{SM}   & 39.49 & 30.16 & 22.93 & 20.01 & 19.37 & 19.29 &
19.21 & 19.31\\
      \textsf{WC}   & 23.98 & 20.58 & 18.20 & 17.03 & \tb{16.59} &
\tb{16.42} & \tb{16.23} & \tb{16.19}\\
      \hline
      $\ \sigma=16$ & 2 & 4 & 8 & 16 & 32 & 64 & 128 & 256\\
      \hline
      \textsf{HOR}  & 25.75 & 19.87 & 17.30 & 16.50 & 16.11 & 15.96 &
15.94 & 16.09\\
      \textsf{QS}   & \tb{20.71} & \tb{18.77} & \tb{16.75} &
\tb{16.29} & \tb{16.00} & \tb{15.90} & 15.92 & 15.98\\
      \textsf{SM}   & 34.79 & 26.22 & 20.39 & 18.10 & 16.89 & 16.66 &
16.52 & 16.83\\
      \textsf{WC}   & 21.08 & 19.01 & 17.06 & 16.39 & 16.07 & 16.02 &
\tb{15.79} & \tb{15.83}\\
      \hline
\end{tabular*}\\[0.05cm]
Running times in hundredths of seconds for \textsf{Rand}$\sigma$
problems
\end{center}
\end{scriptsize}

\begin{scriptsize}
\begin{center}
\begin{tabular*}{0.95\textwidth}{@{\extracolsep{\fill}}|c|cccccccc|}
     \hline
     $\ \sigma=2$ & 2 & 4 & 8 & 16 & 32 & 64 & 128 & 256\\
     \hline
     \textsf{HOR}  & 1.50 & 1.88 & 2.05 & 1.97 & 2.01 & 1.96 & 1.95 & 1.97\\
     \textsf{QS}   & 1.72 & 1.89 & 2.09 & 2.01 & 1.95 & 1.97 & 1.96 & 1.98\\
     \textsf{SM}   & \tb{1.96} & \tb{2.44} & 2.71 & 2.59 & 2.61 &2.56 & 2.56 & 2.59\\
     \textsf{WC}   & 1.72 & 2.18 & \tb{2.75} & \tb{3.16} &\tb{3.66} & \tb{4.09} & \tb{4.60} & \tb{5.20}\\
     \hline
     $\ \sigma=4$ & 2 & 4 & 8 & 16 & 32 & 64 & 128 & 256\\
     \hline
     \textsf{HOR}  & 1.75 & 2.75 & 3.62 & 3.84 & 3.85 & 4.00 & 4.11 & 3.96\\
     \textsf{QS}   & 2.30 & 3.05 & 3.79 & 3.97 & 3.89 & 3.95 & 4.07 & 3.99\\
     \textsf{SM}   & \tb{2.49} & \tb{3.74} & \tb{5.05} & \tb{5.42}& 5.39 & 5.57 & 5.77 & 5.57\\
     \textsf{WC}   & 2.30 & 3.05 & 4.09 & 4.94 & \tb{5.92} &\tb{6.75} & \tb{7.37} & \tb{8.36}\\
     \hline
     $\ \sigma=8$ & 2 & 4 & 8 & 16 & 32 & 64 & 128 & 256\\
     \hline
     \textsf{HOR}  & 1.87 & 3.31 & 5.28 & 7.04 & 7.95 & 8.08 & 8.11 & 8.06\\
     \textsf{QS}   & 2.63 & 3.89 & 5.62 & 7.15 & 7.95 & 8.03 & 8.05 & 8.07\\
     \textsf{SM}   & \tb{2.74} & \tb{4.40} & \tb{7.08} & \tb{9.89}& \tb{11.49} & \tb{11.66} & 11.67 & 11.68\\
     \textsf{WC}   & 2.63 & 3.89 & 5.62 & 7.60 & 9.61 & 11.13 &\tb{12.29} & \tb{13.44}\\
     \hline
     $\ \sigma=8$ & 2 & 4 & 8 & 16 & 32 & 64 & 128 & 256\\
     \hline
     \textsf{HOR}  & 1.93 & 3.63 & 6.46 & 10.24 & 14.10 & 15.91 &16.25 & 15.79\\
     \textsf{QS}   & 2.81 & 4.41 & 7.05 & 10.62 & 14.21 & 15.95 &16.23 & 15.65\\
     \textsf{SM}   & \tb{2.87} & \tb{4.72} & \tb{8.18} & \tb{13.67}& \tb{20.00} & \tb{23.35} & \tb{23.87} & \tb{22.96}\\
     \textsf{WC}   & 2.81 & 4.41 & 7.05 & 10.62 & 14.80 & 18.18 &20.41 & 22.36\\
     \hline
\end{tabular*}\\[0.05cm]
Average advancement for \textsf{Rand}$\sigma$ problems
\end{center}
\end{scriptsize}

\begin{scriptsize}
\begin{center}
\begin{tabular*}{0.95\textwidth}{@{\extracolsep{\fill}}|c|cccccccc|}
      \hline
      $\ \sigma=2$ & 2 & 4 & 8 & 16 & 32 & 64 & 128 & 256\\
      \hline
      \textsf{HOR}  & 48.99 & 65.21 & 99.57 & 126.85 & 139.51 & 137.22
& 137.22 & 133.16\\
      \textsf{QS}   & 45.74 & 61.74 & 94.95 & 123.09 & 135.82 & 134.06
& 124.21 & 120.76\\
      \textsf{SM}   & 86.60 & 123.34 & 149.58 & 186.02 & 205.72 &
201.96 & 205.68 & 201.45\\
      \textsf{WC}   & \tb{43.63} & \tb{60.30} & \tb{90.91} &
\tb{110.29} & \tb{114.81} & \tb{113.07} & \tb{85.10} & \tb{70.05}\\
      \hline
      $\ \sigma=4$ & 2 & 4 & 8 & 16 & 32 & 64 & 128 & 256\\
      \hline
      \textsf{HOR}  & 45.49 & 46.64 & 46.67 & 44.27 & 36.25 & 33.87 &
32.65 & 31.84\\
      \textsf{QS}   & 36.96 & 42.33 & 44.41 & 41.07 & 33.32 & 30.69 &
30.40 & 29.77\\
      \textsf{SM}   & 64.76 & 67.11 & 62.51 & 58.60 & 49.80 & 45.93 &
43.66 & 42.82\\
      \textsf{WC}   & \tb{35.73} & \tb{39.98} & \tb{40.33} &
\tb{34.58} & \tb{27.15} & \tb{24.16} & \tb{22.40} & \tb{21.24}\\
      \hline
      $\ \sigma=8$ & 2 & 4 & 8 & 16 & 32 & 64 & 128 & 256\\
      \hline
      \textsf{HOR}  & 39.09 & 33.88 & 26.99 & 24.10 & 22.18 & 21.39 &
21.25 & 20.54\\
      \textsf{QS}   & \tb{31.50} & \tb{29.99} & 25.70 & 22.80 & 21.68
& 20.63 & 20.62 & 19.83\\
      \textsf{SM}   & 50.46 & 43.41 & 34.09 & 29.50 & 26.62 & 24.94 &
24.64 & 23.13\\
      \textsf{WC}   & 32.36 & 30.12 & \tb{24.97} & \tb{21.30} &
\tb{19.67} & \tb{18.81} & \tb{18.21} & \tb{17.62}\\
      \hline
      $\ \sigma=16$ & 2 & 4 & 8 & 16 & 32 & 64 & 128 & 256\\
      \hline
      \textsf{HOR}  & 33.11 & 24.80 & 20.14 & 18.25 & 17.58 & 17.26 &
17.02 & 16.47\\
      \textsf{QS}   & \tb{25.98} & \tb{22.18} & \tb{19.23} & 17.72 &
17.13 & 16.93 & 16.82 & 16.30\\
      \textsf{SM}   & 42.62 & 33.12 & 25.09 & 21.12 & 19.55 & 18.96 &
18.60 & 17.90\\
      \textsf{WC}   & 26.66 & 22.61 & 19.52 & \tb{17.70} & \tb{17.02}
& \tb{16.66} & \tb{16.40} & \tb{16.02}\\
      \hline
\end{tabular*}\\[0.05cm]
Running times in hundredths of seconds for four
\textsf{Exp}$^5\sigma$ problems
\end{center}
\end{scriptsize}

\begin{scriptsize}
\begin{center}
\begin{tabular*}{0.95\textwidth}{@{\extracolsep{\fill}}|c|cccccccc|}
     \hline
     $\ \sigma=2$ & 2 & 4 & 8 & 16 & 32 & 64 & 128 & 256\\
     \hline
     \textsf{HOR}  & 1.04 & 1.11 & 1.23 & 1.41 & 1.63 & 1.87 & 1.86 & 1.97\\
     \textsf{QS}   & 1.08 & 1.14 & 1.24 & 1.40 & 1.64 & 1.86 & 1.85 & 1.97\\
     \textsf{SM}   & \tb{1.10} & 1.17 & 1.29 & 1.49 & 1.72 & 1.97 & 1.98 & 2.06\\
     \textsf{WC}   & \tb{1.10} & \tb{1.21} & \tb{1.41} & \tb{1.67}& \tb{2.02} & \tb{2.34} & \tb{2.90} & \tb{3.55}\\
     \hline
     $\ \sigma=4$ & 2 & 4 & 8 & 16 & 32 & 64 & 128 & 256\\
     \hline
     \textsf{HOR}  & 1.32 & 1.65 & 2.04 & 2.24 & 2.53 & 2.81 & 3.08 & 3.17\\
     \textsf{QS}   & 1.54 & 1.74 & 2.10 & 2.35 & 2.53 & 2.87 & 3.06 & 3.12\\
     \textsf{SM}   & \tb{1.68} & \tb{2.03} & \tb{2.60} & 2.88 &3.29 & 3.70 & 4.13 & 4.28\\
     \textsf{WC}   & 1.62 & 1.98 & 2.45 & \tb{3.09} & \tb{3.80} &\tb{4.60} & \tb{5.59} & \tb{6.34}\\
     \hline
     $\ \sigma=8$ & 2 & 4 & 8 & 16 & 32 & 64 & 128 & 256\\
     \hline
     \textsf{HOR}  & 1.63 & 2.29 & 3.10 & 3.71 & 4.49 & 4.99 & 5.24 & 5.74\\
     \textsf{QS}   & 2.02 & 2.59 & 3.20 & 3.77 & 4.38 & 4.98 & 5.15 & 5.80\\
     \textsf{SM}   & \tb{2.25} & \tb{3.17} & \tb{4.34} & \tb{5.36}& \tb{6.65} & \tb{7.59} & 8.08 & 9.05\\
     \textsf{WC}   & 2.04 & 2.70 & 3.58 & 4.68 & 5.91 & 6.95 &\tb{8.24} & \tb{9.66}\\
     \hline
     $\ \sigma=16$ & 2 & 4 & 8 & 16 & 32 & 64 & 128 & 256\\
     \hline
     \textsf{HOR}  & 1.79 & 2.99 & 4.46 & 5.97 & 7.29 & 8.26 & 9.65 & 10.25\\
     \textsf{QS}   & 1.79 & 2.99 & 4.46 & 5.97 & 7.29 & 8.26 & 9.65 & 10.25\\
     \textsf{SM}   & \tb{2.61} & \tb{4.07} & \tb{6.17} & \tb{8.69}& \tb{11.08} & \tb{13.03} & \tb{15.44} & \tb{16.76}\\
     \textsf{WC}   & 2.46 & 3.49 & 4.87 & 6.72 & 8.57 & 10.55 &12.83 & 14.95\\
     \hline
\end{tabular*}\\[0.05cm]
Average advancements for four \textsf{Exp}$^5\sigma$ problems
\end{center}
\end{scriptsize}

The above experimental results show that the algorithm based on the
worst-character heuristic obtains the best runtime performances in
most cases, especially for long patterns and small alphabets, and it
is second only to the \quick algorithm, in the case of small patterns,
as the alphabet size increases.


 Concerning the average advancements, it turns out that the proposed
 heuristic is quite close to the \smith heuristic, which generally
 shows the best behavior.  We notice, though, that in the case of
 long patterns and small alphabets the presented heuristic proposes
 the longest average advancements.

Finally we observe that the performances of the worst-character
heuristic
increase when tested on an \textsf{Exp}$^5\sigma$ problem.

\newpage

\section{Conclusions}\label{sec:conclusions}
%

Several efficient variations of the bad-character heuristic have been
proposed in the last years with the aim of obtaining better
performances in practical cases.  For instance, the \berry
algorithm~\cite{BR99} generalizes the \quick algorithm by using in its
bad-character rule the last two characters, rather than just the last
one.  Another example is the \tbm algorithm \cite{HS91} which
introduces, using the Horspool bad-character rule, an efficient
implementation of the searching phase.  Finally, algorithms in the \fs
family~\cite{CF05} combine the bad-character rule with the good-suffix
heuristic by computing an $\mathcal{O}(\sigma \times m)$-space
function.

In this paper we have presented the \emph{worst-character rule}, a
variation of the \emph{bad-character} heuristic, which is based on the
position relative to the current shift which yields the largest
average advancement, according to the characters distribution in the
text.  We have also shown experimental evidence that the
worst-character rule achieves very good results in practice,
especially in the case of long patterns or small alphabets in random
texts and in the case of texts in natural languages.

%

\bibliographystyle{alpha}

\end{document}